\documentclass{article}
\usepackage{authblk}
\usepackage{spconf}
\usepackage{amsmath}
\usepackage{epsfig}
\usepackage{amssymb}
\usepackage{footmisc}
\usepackage{multirow}
\usepackage{algorithm}     
\usepackage{algorithmic}
\usepackage{setspace}
\usepackage{booktabs} 
\usepackage[bookmarks=false]{hyperref}
\usepackage{graphicx}
\usepackage[numbers,sort&compress]{natbib}
\usepackage{textcomp}
\begin{document}\sloppy

\def\x{{\mathbf x}}
\def\L{{\cal L}}

\title{Semi-supervised Compatibility Learning Across Categories for Clothing Matching}



 \name{Zekun Li$^{1,3}$, Zeyu Cui$^{2,3}$, Shu Wu$^{2,3,4}$, Xiaoyu Zhang$^{1}$ and Liang Wang$^{2,3}$ \thanks{The first two authors Zekun Li and Zeyu Cui are listed as joint first authors. Shu Wu and Xiaoyu Zhang are both corresponding authors.
}}
                 \address{ $^{1}$Institute of Information Engineering, Chinese Academy of Sciences\\
                          $^{2}$Institute of Automation, Chinese Academy of Sciences \\
                          $^{3}$University of Chinese Academy of Sciences \\
                          $^{4}$Jiaozhou Artificial Intelligence Research, Chinese Academy of Sciences \\
                         lizekunlee@gmail.com,
						\{zeyu.cui,shu.wu\}@nlpr.ia.ac.cn, \\ 
                          zhangxiaoyu@iie.ac.cn,
                          	wangliang@nlpr.ia.ac.cn}

\maketitle

\begin{abstract}
Learning the compatibility between fashion items across categories is a key task in fashion analysis, which can decode the secret of clothing matching. 
The main idea of this task is to map items into a latent style space where compatible items stay close.
Previous works try to build such a transformation by minimizing the distances between annotated compatible items, which require massive  item-level supervision.
However, these annotated data are expensive to obtain and hard to cover the numerous items with various styles in real applications. In such cases, these supervised methods fail to achieve satisfactory performances.
In this work, we propose a semi-supervised method to learn the compatibility across categories.
We observe that the distributions of different categories have intrinsic similar structures.
Accordingly, the better distributions align, the closer compatible items across these categories become. 
To achieve the alignment, we minimize the distances 
between distributions with unsupervised adversarial learning,
and also the distances between some annotated compatible items which play the role of \emph{anchor points} to help align.
Experimental results on two real-world datasets demonstrate the effectiveness of our method. 

\end{abstract}

\begin{keywords}
Semi-supervised compatibility learning, Clothing matching, Adversarial learning
\end{keywords}

\section{Introduction}
Nowadays, clothes with various styles are increasing quickly, broadening the range of people's choices.
``Which pair of shoes should I select to match the jeans?'', such a problem has become a daily headache for many people.  
Solving this problem requires learning the compatibility between fashion items across categories. 
As a matter of fact, 
many existing efforts have been dedicated to the task of compatibility learning. 
The key idea is to map the items into a 
latent style space where compatible items would stay close. 
Previous works 
\cite{Mcauley2015Image,He2017Learning}
 try to learn the transformation by minimizing the distance between annotated compatible items in the style space, which requires massive supervision to be general. 
However, these annotated data are expensive to obtain and hard to cover the numerous and increasing clothing items in real applications, and so these supervised methods often fail to achieve satisfactory performance. 
How to learn such a general transformation for numerous items of various styles with a limited amount of supervision has become a demanding problem.

\begin{figure}[t]
\includegraphics[width=1\linewidth]{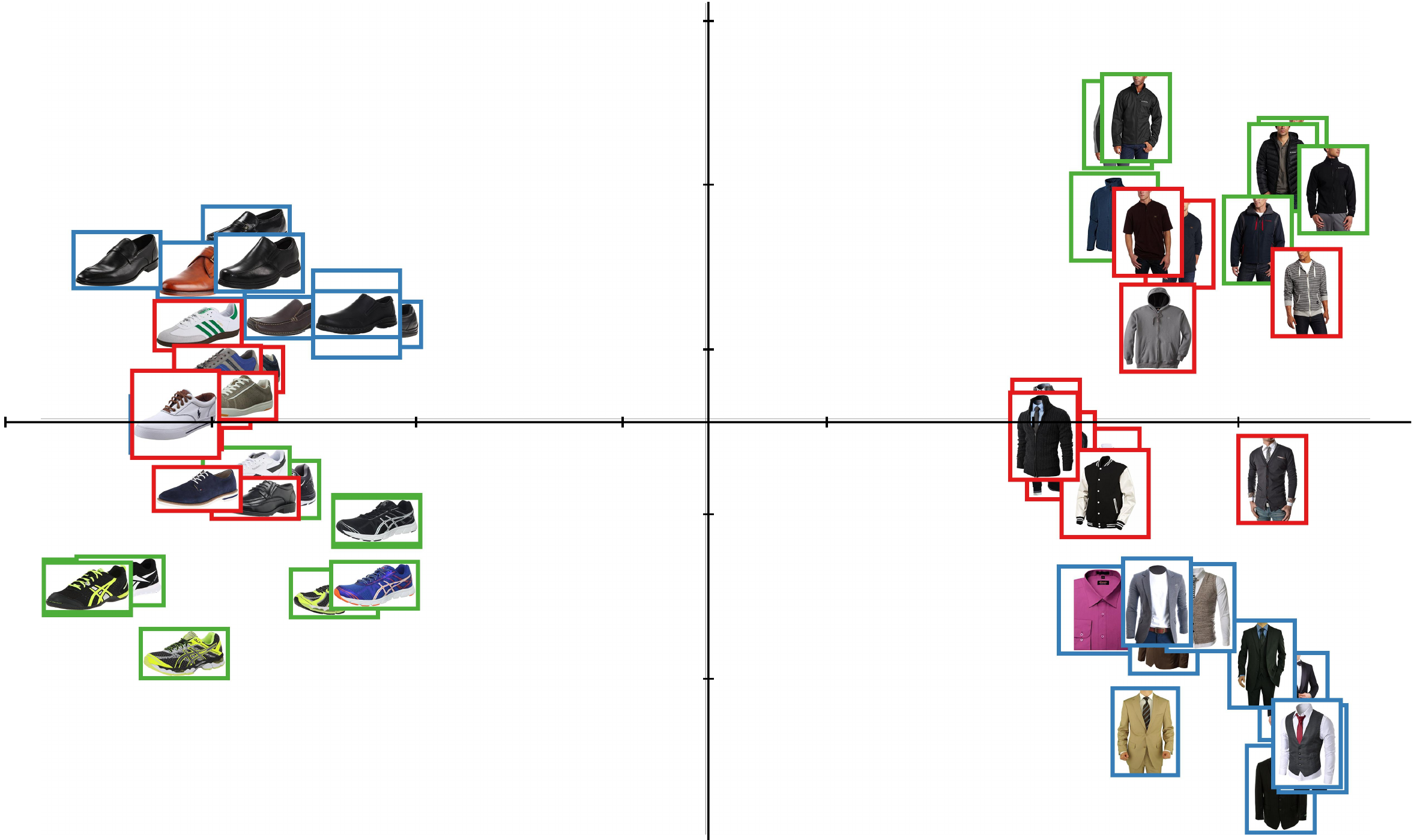}
\vspace{-8mm}
\caption{
2-D visualization of the distributions of tops and shoes, which have similar structures despite different orientations and scales. 
The tops and shoes in frames with the same color have same styles and are compatible.
}
\vspace{-5mm}
\label{fig:pca}
\end{figure}

The key to this task is to make compatible items close in the learned style space.
Previous works consider it only at the item level (i.e., minimize the distances between annotated compatible items), which inevitably require enough item-level supervision to be general. 
In fact, some information at a higher level is ignored.
Here we look into the distribution level.
We randomly select some tops and shoes and visualize their distributions respectively in Figure \ref{fig:pca}, with raw image features reduced to 2-D vectors by Principal Component Analysis (PCA). 
Different colors of frames represent different styles.
The tops and shoes in frames of the same color have same styles and are compatible. 
As can be seen, in spite of different orientations and scales, these distributions have intrinsic similar structures according to styles.
Accordingly, there may exist a desired transformation to align the distributions of different categories in the style space, which can make compatible items with same styles close in return.


In fact, 
the method of distribution alignment 
has be proven to be effective on the bilingual lexicon induction task 
in the domain of natural language processing (NLP), which is highly analogous to our task.
The studies on the bilingual lexicon induction task follow the idea to map words into a word embedding space.
There is evidence that different languages represent semantic concepts with similar structure leading to the structural isomorphism across word embedding spaces of different languages \cite{Youn2016On}.
Viewing word embedding spaces as distributions, Zhang \emph{et al.} proposed to build the cross-lingual connection by minimizing their earth mover’s distance \cite{Zhang2017Earth}.
Differently, they achieve the alignment only by minimizing the distance between distributions while we find the combination of distribution-level and item-level distance minimization can better align the distributions. 

In this paper, we propose Semi-supervised Compatibility learning Generative Adversarial Networks (\textbf{SC-GANs}) to learn the compatibility across categories, not requiring massive supervision.
In our semi-supervised method, the unsupervised distribution-level distance minimization is combined with the supervised item-level one to build a general transformation, which can align the distributions of different categories and make compatible items with same styles close in the style space.
An adversarial learning strategy is adopted to minimize both the Wasserstein distance between distributions and Euclidean distance between compatible items.
We conduct experiments to evaluate the performance of our method on two real-world datasets. 
Our semi-supervised method shows superiority over other supervised methods when lacking massive supervision, which is effective in real applications.
The code and data has been released\footnote{\url{https://github.com/CRIPAC-DIG/SCGAN}}.

Our main contributions can be summarized in threefold:

\begin{itemize}
\setlength{\itemsep}{2pt}
\setlength{\parsep}{0pt}
\setlength{\parskip}{2pt}
\item We first propose that aligning distributions of different categories in the style space can make compatible items with same styles close, which can be achieved by 
minimizing the distances between distributions and also some annotated compatible items.
\item We propose a semi-supervised method \textbf{SC-GANs} to learn compatibility across categories not requiring massive supervision, which is potential in real applications.
\item Experimental results on two real-world datasets demonstrate the effectiveness of our proposed method.
\end{itemize}



\section{The Model}


The whole item set is denoted as $\mathcal{I}$ while the set of category is $\mathcal{C} = \{ c_{1},c_{2}, ... \}$. 
The set of items in category $c_{i}$ is $\mathcal{I}_{c_{i}}$.
The set of compatible pairs is $\mathcal{P}$.
The compatibility of two items $x \in \mathcal{I}_{c_{j}}, y \in \mathcal{I}_{c_{j}}$, is $r(x, y)$. 
Our goal is to estimate the value of $r(x, y)$. 
$\mathbf{v}_{x}$ is a high-dimensional feature vector of item $x$ extracted from its image. 
The distribution of category $c$ in the style space is 
$\mathbb{P}_{c}$.
We aim to find the exact style transformation to map items (their feature vectors) of different categories into one style space, where the distributions of different categories align and the compatible items are close as well.
In the style space, the distance between items $x$ and $y$ is $d(x, y)$, which can indicate $r(x, y)$.
The lower $d(x, y)$, the higher $r(x, y)$ is, and the more compatible $x$ and $y$ are.

\subsection{Preliminaries}

\noindent\textbf{Feature Extraction.} 
\label{sect:feature_extractor}
The visual features of items are extracted from their images using deep convolution networks, VGG-16 \cite{Simonyan2014Very}, which is widely used for image representation learning \cite{Mcauley2015Image,Veit2015Learning,He2017Learning}.
It has been pre-trained on large-scale ImageNet images. We adopt the output of the second fully connected layer, a 4096-dimensional feature vector. 

\noindent\textbf{Style Space Transformation.}
\label{sect:stylespace}
Compatible items usually have similar styles. Previous works
assume that there exists a style space where compatible items stay close.
Veit et al.~\cite{Veit2015Learning} use Siamese CNNs to learn a feature transformation from the image space to the style space.
McAuley et al.~\cite{Mcauley2015Image} use Low-rank Mahalanobis Transformation (LMT) to map compatible items to close positions in the style space.
He et al.~\cite{He2017Learning} map items into several style spaces to compute a weighted sum of the $K$ distances between two items, which can deal with diversity across different query items. 
Liu et al.~\cite{liu2017deepstyle} proposed to map items into a style space where the categorical information are eliminated.
The above methods build such a transformation only by minimizing the distance between annotated compatible items.
Nevertheless, we build such a transformation by aligning the distributions, i.e., minimizing the distance between the distributions of different categories as well as annotated compatible items.

\noindent \textbf{Wasserstein Distance.}
In this work, we adopt the Wasserstein distance as the measure of distance between distributions.
Wasserstein distance is a measure of distance between probability distributions, which can be formulated as,
\begin{equation}\label{equ:wassdist}
W(\mathbb{P}_1,\mathbb{P}_2)=\inf_{\gamma \in \Gamma (\mathbb{P}_1,\mathbb{P}_2)}\mathbb{E}_{(x,y) \sim \gamma }[c(x,y)],
\end{equation}
where $\Gamma (\mathbb{P}_1,\mathbb{P}_2)$ denotes the set of all joint distributions $\gamma(x,y)$ with marginals $\mathbb{P}_1$
and $\mathbb{P}_2$.
It can be considered as the continuous case of the Earth Mover's Distance, a powerful tool widely used in computer vision and natural language processing \cite{Rubner1998Metric, Zhang2017Earth}.
Intuitively, if each distribution is viewed as a unit amount of ``dirt", earth mover's distance is the minimum ``cost" of turning one pile into the other, which is assumed to be the amount of dirt that needs to be moved multiplies the distance it has to be moved. This conforms to the nature of our task, i.e., put an item close to its compatible item of another category in the latent style space.
In order to minimize the Wasserstein distance between distributions, we adopt an adversarial learning strategy similar with WGANs.


\noindent \textbf{Wasserstein GANs.}
Goodfellow et al.~\cite{Goodfellow2014Generative} first propose GANs to generate distribution similar with the target distribution. But the original GANs are difficult to train.  Many efforts have been devoted to solve the problem. Arjovsky \emph{et al.} propose WGANs~\cite{Arjovsky2017Wasserstein} to deal with this training problem. WGANs can be viewed as an adversarial game to minimize the Wasserstein distance between the generated distribution $\mathbb{P}_{g}$ and the real distribution $\mathbb{P}_{r}$.
With the ground distance $c$ being the Euclidean distance L2, Eq.(\ref{equ:wassdist}) can be cast to the following equation according to the Kantorovich-Rubinstein duality,
\begin{equation}\label{equ:wassdist2}
W(\mathbb{P}_{g},\mathbb{P}_{r}) = \frac{1}{K}\sup_{\left \| f \right \|_{L}\leq{K}}\mathbb{E}_{x\sim{\mathbb{P}_{g}}}[f(x)]-\mathbb{E}_{x\sim{\mathbb{P}_{r}}}[f(x)],
\end{equation}
where the supremum is over all $K$-Lipschitz functions $f$.

WGANs consist of two components, critic $D$ and generator $G$.
The critic $D$ is a neural network to approximate $f$ in Eq.(\ref{equ:wassdist2}) with weight clipping to ensure that the function family is $K$-Lipschitz.
It manages to distinguish the critic real distribution and the generated distribution, so as to maximize their Wasserstein distance. The objective of $D$ is

\begin{equation}\label{equ:traind}
\max_{D} \mathbb{E}_{x\sim{\mathbb{P}_{r}}}[D(x)]-\mathbb{E}_{x\sim{\mathbb{P}_{g}}}[D(x)].
\end{equation}
When the objective (\ref{equ:traind})  is trained until optimality, it approximates the Wasserstein distance.
The generator $G$ then aims to minimize the approximate Wasserstein distance, which leads to
\begin{equation}\label{equ:traing}
\min_{G} -\mathbb{E}_{x\sim{\mathbb{P}_{g}}}[D(x)]
\end{equation}

\begin{figure}[t]
\includegraphics[width=1\linewidth]{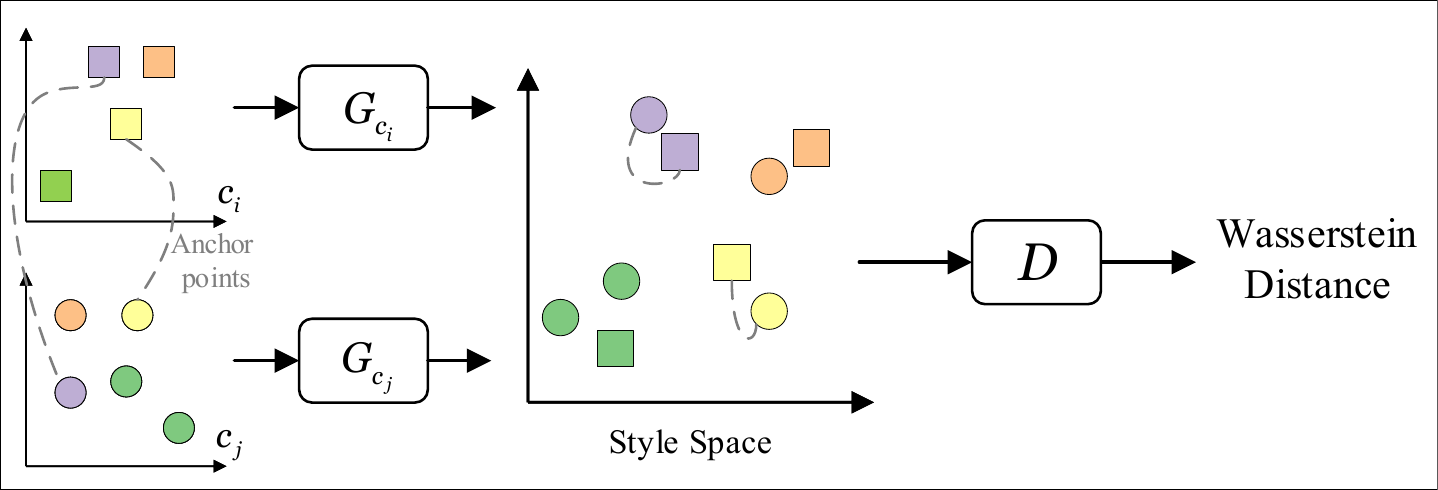}
\vspace{-5mm}
\caption{
The framework of SC-GANs. Items of each category (we only show two categories here briefly) are mapped into one style space with category-specific generators.
The critic $D$ estimates the Wasserstein distance, which will be passed to the generators and guide them towards minimizing the Wasserstein estimate. The generators also try to minimize the distance between annotated compatible items, which play the role of \emph{anchor points} to help align.
}
\vspace{-4mm}
\label{fig:framework}
\end{figure}

\subsection{Model Architecture}
The framework of our model is shown in Figure \ref{fig:framework}.
Our model consists of two components, generator and critic.
The part of generator plays the role of style space transformation to map items into the latent style space.
The part of critic estimates the Wasserstein distance between the transformed distributions. 
In our model, we have one critic $D$ and category-specific generators $G=\{\mathbf{G}_{c_{1}}, \mathbf{G}_{c_{2}}, ...\}$, since each category has its own unique characteristics.
Accordingly, the style vector $\mathbf{s}_{x}$ of item $x$ in the style space can be calculated as,
\begin{equation} \label{eqh_style}
	 \mathbf{s}_{x}
	  = \mathbf{G}_{c_{i}} \mathbf{v}_{x},
\end{equation}
where $x \in \mathcal{I}_{c_{i}}$, $\mathbf{G}_{c_{i}}$ is the transformation matrix corresponding to the category $c_{i}$.

\noindent \textbf{Distribution-level Distance Minimization.}
This is an \textbf{unsupervised} part, not requiring supervision of compatible pairs.
Different from WGANs minimizing the Wasserstein distance between the generated distribution and real distribution,
SC-GANs minimize the Wasserstein distance between dyadic transformed category distributions in the style space as shown in Figure \ref{fig:framework}. 
We here take two categories $\emph{c}_{i}$, $\emph{c}_{j}$ for example.
The Wasserstein distance between transformed distribution in the latent style space $\mathbb{P}_{{c_{i}}}$ and $\mathbb{P}_{{c_{j}}}$ of categories $\emph{c}_{i}$, $\emph{c}_{j}$ can be estimated as:

\begin{equation}\label{eq5}
\begin{aligned}
&W(\mathbb{P}_{{c_{i}}},\mathbb{P}_{{c_{j}}}) = \\
&\frac{1}{K}\sup_{\left \| f \right \|_{L}\leq{K}}
\mathbb{E}_{x\sim{\mathcal{I}_{c_{i}}}}[f(\mathbf{G}_{c_{i}} \mathbf{v}_{x})]- 
\mathbb{E}_{x\sim{\mathcal{I}_{c_{j}}}}[f(\mathbf{G}_{c_{j}}\mathbf{v}_{x})].
\end{aligned}
\end{equation}
Following WGANs, we approximate $f$ with the critic $D$, whose objective is,
\begin{equation}\label{equ:uclansd}
\max_{D} \mathbb{E}_{x\sim{\mathcal{I}_{c_{i}}}}[D(\mathbf{G}_{c_{i}}\mathbf{v}_{x})]-\mathbb{E}_{x\sim{\mathcal{I}_{c_{j}}}}[D(\mathbf{G}_{c_{j}}\mathbf{v}_{x})].
\end{equation}
When the objective (\ref{equ:uclansd}) is trained until optimality, it approximates the Wasserstein distance between distributions $\mathbb{P}_{{c_{i}}}$ and $\mathbb{P}_{{c_{j}}}$. The generators aim to minimize this distance as,
\begin{equation}\label{equ:uclansg}
\min_{\mathbf{G}_{c_{i}},\mathbf{G}_{c_{j}}} \mathbb{E}_{x\sim{\mathcal{I}_{c_{i}}}}[{D}(\mathbf{G}_{c_{i}}\mathbf{v}_{x})] - \mathbb{E}_{x\sim{\mathcal{I}_{c_{j}}}}[D(\mathbf{G}_{c_{j}}\mathbf{v}_{x})].
\end{equation}

\noindent \textbf{Item-level Distance Minimization.}
This is a \textbf{supervised} part.
The generators should also keep the compatible pairs close, which play the role of \emph{anchor points} to help align the distributions.
In addition, we impose an orthogonal constraint on the transformation matrices to keep structural information, according to \cite{Zhang2017Earth}. 
Overall, we have the following objective for generators:
\begin{equation}\label{equ:uclansg2}
\begin{aligned}
&\min_{\mathbf{G}_{c_{i}},\mathbf{G}_{c_{j}}} \mathbb{E}_{x\sim{\mathcal{I}_{c_{i}}}}[{D}(\mathbf{G}_{c_{i}}\mathbf{v}_{x})] - \mathbb{E}_{x\sim{\mathcal{I}_{c_{j}}}}[D(\mathbf{G}_{c_{j}}\mathbf{v}_{x})] \\
& + \eta \sum_{x \in \mathcal{I}_{c_{i}}, y \in \mathcal{I}_{c_{j}}, (x,y) \in \mathcal{P}} \left \| \mathbf{G}_{c_{i}} \mathbf{v}_{x} - \mathbf{G}_{c_{j}} \mathbf{v}_{y} \right \|_{2}^{2} \\
& + \lambda \left \| \mathbf{G}_{c_{i}}\mathbf{G}_{c_{i}}^{T}- \mathbf{E} \right \|_{F} + \lambda \left \| \mathbf{G}_{c_{j}}\mathbf{G}_{c_{j}}^{T}- \mathbf{E} \right \|_{F}
,
\end{aligned}
\end{equation}
where $\lambda$, $\eta$ are coefficients and $\mathbf{E}$ is the identity matrix.

\begin{algorithm}[t]
\setstretch{0.9}
    \caption{SC-GANs}
    
	\label{alg:learnstrategy}
    \begin{algorithmic}[1]%
    \REQUIRE $G=\{\mathbf{G}_{c_{1}}, \mathbf{G}_{c_{2}}, ...\}$: generators. $D$: critic. $m$: batch size. $l$: the gradient clip bound. $\lambda$, $\eta$: coefficients. 
    $n_{critic}$: the number of critic iterations per generator iteration. 
    
    	\STATE Randomly initialize $G$ and  $D$, set $t=0$;
    	\WHILE {$(G,D)$ not converged}
    		\STATE Sample $c_{i}$, $c_{j} \in \mathcal{C}$ ; $t \leftarrow  t +1$;
    		\STATE Sample $\{x^{(i)}\}^{m}_{i=1} \in \mathcal{I}_{c_{i}}$, a batch of items from $\mathcal{I}_{c_{i}}$;
    		\STATE Sample $\{y^{(i)}\}^{m}_{i=1} \in \mathcal{I}_{c_{j}}$, a batch of items from $\mathcal{I}_{c_{j}}$;
    		\STATE Sample $(x,y) \in \mathcal{P}, x \in \mathcal{I}_{c_{i}}, y \in \mathcal{I}_{c_{j}}$, a batch of compatible pairs from $\mathcal{P}$;
    		\IF{$t \mod n_{critic} = 0$}
    			\STATE Update $\mathbf{G}_{c_{i}}$ and $\mathbf{G}_{c_{j}}$ 
    			by descending:
    			
    			$\frac{1}{m} \sum^{m}_{i=1} {D}(\mathbf{G}_{c_{i}}x_{i}) - \frac{1}{m} \sum^{m}_{i=1} {D}(\mathbf{G}_{c_{j}}y_{i})$
    			
    			$+ \eta \sum_{x \in \mathcal{I}_{c_{i}}, y \in \mathcal{I}_{c_{j}}, (x,y) \in \mathcal{P}} \left \| \mathbf{G}_{c_{i}} \mathbf{v}_{x} - \mathbf{G}_{c_{j}} \mathbf{v}_{y} \right \|_{2}^{2}$
    			
    			$+ \lambda \left \| \mathbf{G}_{c_{i}}\mathbf{G}_{c_{i}}^{T} - \mathbf{E} \right \|_{F} + \lambda \left \| \mathbf{G}_{c_{j}}\mathbf{G}_{c_{j}}^{T} - \mathbf{E} \right \|_{F}$;
    			
    		\ENDIF
    			\STATE Update $D$ 
    			by ascending:
    			
    			$\frac{1}{m} \sum^{m}_{i=1} {D}(\mathbf{G}_{c_{1}}x_{i}) - \frac{1}{m} \sum^{m}_{i=1} {D}(\mathbf{G}_{c_{2}}y_{i})$;
    			\STATE $D \leftarrow clip(D, -l, l)$;

    		
    	\ENDWHILE
  	
    \end{algorithmic}

\end{algorithm}


%
%
%

\noindent \textbf{Learning the Model.}
The training process of our model is shown in Algorithm \ref{alg:learnstrategy}. 
Since there are many categories in the training dataset, we only train the critic and two randomly selected category-specific generators each time. 
The generators and the critic are trained in a settled proportion $n_{critic}$ (i.e., train the critic $n_{critic}$ iterations per generator iteration), until the critic and all the generators converge.

\noindent \textbf{Distance and Compatibility.}
After training, the compatible items of different categories are close in the learned style space.
Therefore, we can calculate the distance between items $x \in \mathcal{I}_{c_{i}}$, $y \in \mathcal{I}_{c_{j}}$ in the style space as,
\begin{equation}\label{equ:compatibility}
d(x,y)=  \left \| \mathbf{G}_{c_{i}}\mathbf{v}_{x}-\mathbf{G}_{c_{j}}\mathbf{v}_{y} \right \|_{2}^{2}.
\end{equation}
Following \cite{Mcauley2015Image}, 
their compatibility is related to the distance as, 
\begin{equation}\label{equ:compatibility}
r(x,y)=  \sigma(-d(x,y))=\frac{1}{1+e^{d(x,y)}}.
\end{equation}

\section{Experiment}
\subsection{Datasets}
We conduct experiments on two datasets: the Amazon ``also-bought'' dataset and the Taobao dataset.   

\noindent \textbf{Amazon ``also-bought'' dataset.}
Amazon dataset was collected by McAuley et al.~\cite{Mcauley2015Image}.
Following previous work, the ``also-bought'' relationships are used as compatibility in the five clothing categories, ``Women", ``Men", ``Girls", ``Boys" and ``Baby".
There are totally 1101118 items from 263 subcategories and 3457219 relationships covering all the items.
Although it is commonly used by previous works \cite{Veit2015Learning,Mcauley2015Image,He2017Learning}, the relationship ``also-bought'' is not totally equal to compatibility.
To compare our method with others 
in case of lacking enough supervision covering all the items, 
we randomly select 0.5, 1, 2 permillage of compatible pairs in the Amazon dataset as seeds to form the training dataset
and $20\%$ to form the testing set.

\noindent \textbf{Taobao dataset.}
Taobao dataset is a collection of outfits of women clothing on Taobao.com released by Alibaba Group\footnote{
\url{https://tianchi.aliyun.com/datalab/index.html}.}, in which compatibility was manually labelled by fashion experts.
The taobao dataset consists of 499983 items from 71 categories.
There are 407152 compatibility relationships covering 60767 items.
Since the compatibility doesn't cover all the items, it's suitable to test the performances of these methods in the real scenario.
We thus conduct experiments on the \emph{whole} dataset with $80\%$ for training and $20\%$ for testing.
For the two datasets (Amazon and Tabao), we denote the training set as $\mathcal{P}_{train}$ and testing set as $\mathcal{P}_{test}$.

\subsection{Compared Methods}

\textbf{Nearest Neighborhood (NN)} is a traditional unsupervised method. The dimensionality of raw item features is reduced to $d$ by PCA, and then their Euclidean distance are used to measure the compatibility.
\textbf{Category Tree (CT)} measures the compatibility between two items using the co-occurrences between their categories.
\textbf{Low-rank Mahalanobis Transform (LMT)} models the relationships between items in the style space via a single low-rank Mahalanobis embedding matrix \cite{Mcauley2015Image}.
\textbf{Mixtures of Non-Metric Embeddings for Recommendation (Monomer)}
is proposed by He~\cite{He2017Learning}.  This method maps the compatible items into $K$ latent style spaces to compute weighted sum of the $K$ distances between the two items.
\textbf{UC-GANs} is the unsupervised version of \textbf{SC-GANs}, with only distribution-level distance minimization.
\subsection{Experimental Settings}
We set the dimensionality of style vectors $d$ in all compared methods as $128$, the orthogonal regularization coefficient $\lambda$ as $0.01$, $\eta$ as $0.1$. 
RMSProp 
is adopted for gradient descent, with the learning rate $0.001$. 
The gradient clip bound $l$ is $-1$ and the batch size $m$ is $30$. 
$n_{critic}$ is 5, i.e., we train the critic $5$ iterations per generator iteration. 
When testing, for each compatible pair $(x, y) \in \mathcal{P}_{test}$ we randomly select an item to replace $y$ to generate a negative pair $(x, y^{-})$. 
We adopt the widely used AUC (Area Under the ROC curve) as metric,
\begin{equation}\label{eq1}
AUC = \frac{1}{\left | \mathcal{P}_{test} \right |}\sum_{\left ( x, y \right ) \in \mathcal{P}_{test}} \delta \left ( r(x, y) > r(x, y^{-}) \right ),
\end{equation}
where $\delta \left( a \right )$ is an indicator function
that returns one if the argument $a$ is \emph{true} and zero otherwise.


\begin{table}[htbp]
\vspace{-4mm}
  \centering
  \caption{Performance comparison on the Amazon ``also-bought'' dataset evaluated by AUC. Seed refers to the permillage of compatible pairs in the whole dataset.}
\vspace{2mm}
\scalebox{0.815}{
    \begin{tabular}{ccccccc}
    \toprule
    Method & Seed  & Women & Men   & Girls & Boys  & Baby \\
    \midrule
    NN    & 0     & 0.5674  & 0.5900  & 0.5229  & 0.5799  & 0.5340  \\
    \midrule
    \multirow{3}[2]{*}{CT} & 0.5   & 0.5813  & 0.5943  & 0.5100  & 0.5181  & 0.5206  \\
          & 1     & 0.6159  & 0.6235  & 0.5510  & 0.5796  & 0.5859  \\
          & 2     & 0.6373  & 0.6341  & 0.6073  & 0.6182  & 0.6049  \\
    \midrule
    \multirow{3}[2]{*}{LMT} & 0.5   & 0.6802  & 0.6722  & 0.6162  & 0.6062  & 0.6301  \\
          & 1     & 0.6892  & 0.6812  & 0.6616  & 0.6822  & 0.6518  \\
          & 2     & 0.7270  & 0.7111  & 0.6707  & 0.6984  & 0.7391  \\
    \midrule
    \multirow{3}[2]{*}{Monomer} & 0.5   & 0.6897  & 0.6892  & 0.6613  & 0.6691  & 0.6307  \\
          & 1     & 0.6911  & 0.6923  & 0.6977  & 0.6742  & 0.6422  \\
          & 2     & 0.7311  & 0.7403  & 0.7210  & 0.7301  & 0.6506  \\
    \midrule
	UC-GANs & 0     & 0.7369  & 0.7248  & 0.6916  & 0.7307  & 0.6565  \\
	\midrule
    \multirow{3}[2]{*}{SC-GANs} 
          & 0.5   & 0.7554  & 0.7620  & 0.7183  & 0.7311  & 0.6997  \\
          & 1     & 0.7634  & 0.7702  & 0.7682  & 0.7391  & 0.7297  \\
          & 2     & \textbf{0.7899 } & \textbf{0.7906 } & \textbf{0.7778 } & \textbf{0.7574 } & \textbf{0.7434 } \\
    \bottomrule
    \end{tabular}}
   \label{tab:performance}
\end{table}%


\begin{table}[htbp]
\vspace{-2mm}
  \centering
  \caption{Performance comparison evaluated by AUC on the \emph{whole} Taobao Dataset, in which the annotated compatibility relationships don't cover all the items.}
\vspace{2mm}
\scalebox{0.755}{
    \begin{tabular}{ccccccc}
    \toprule
    Method & NN    & CT    & LMT   & Monomer & UC-GANs & SC-GANs \\
    \midrule
    TaoBao & 0.5183  & 0.6168  & 0.7335  & 0.7897  & 0.8239  & \textbf{0.8421 } \\
    \bottomrule
    \end{tabular}}
\label{tab:taobao}
\end{table}%

\subsection{Performance Comparison}

The performances on Amazon dataset and Taobao dataset
are shown in Table \ref{tab:performance} and \ref{tab:taobao} respectively.
On both datasets, CT achieves better performance than NN, which indicates that categorical information is necessary for learning compatibility.
As can be seen, the performances of supervised models improve with supervision increasing on Amazon dataset, which demonstrates that these supervised methods need massive supervision to achieve considerable performance.
Compared with the supervised methods LMT and Monomer, 
UC-GANs achieve highly competitive performance on Amazon dataset and much better performance on 
Taobao dataset, which may due to the higher quality of annotated compatible relations in Taobao dataset. 
In a word, this finding confirms the effectiveness of distribution alignment on the task of compatibility learning across categories. 
SC-GANs outperform UC-GANs with item-level supervision, suggesting the combination of item-level distance minimization and distribution-level can better align distributions indeed. 
Overall, we can see that our semi-supervised method SC-GANs outperform other supervised methods in case of lacking massive supervision covering enough items, which is effective in real applications.

\subsection{Model Analysis}

\begin{figure}[t]
\centering
\includegraphics[width=0.8\linewidth]{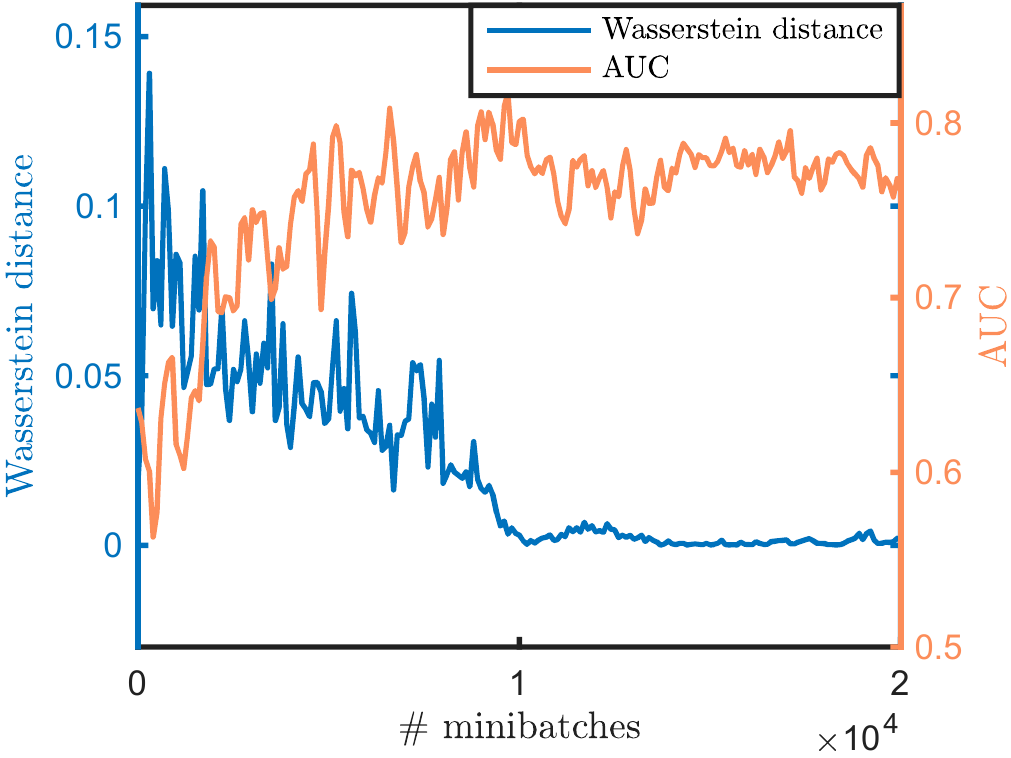}
\vspace{-2mm}
\caption{
The change of AUC and Wasserstein distance along with the training process on a toy dataset consisting of two categories in the Taobao dataset.}
\vspace{-3mm}
\label{fig:auc_wass}
\end{figure}
We first verify that aligning the distributions can make the compatible items close.
The Wasserstein distance between two distributions indicates the degree of their alignment.
We train the model on a toy dataset consisting of two randomly selected categories in the Taobao dataset.    
The change of AUC and Wasserstein distance along with the training process is shown in Figure \ref{fig:auc_wass}. 
It is obvious that the Wasserstein distance is strongly correlated with AUC, which suggests that it's effective to minimize the Wasserstein distance (aligning the distributions) for shortening the distances between compatible items.

\begin{table}[t]\footnotesize
\centering
\caption{The AUC results of ablation study on SC-GANs}
\vspace{2mm}
\scalebox{0.88}{
    \begin{tabular}{ccccccc}
    \toprule
    \multirow{2}[4]{*}{Method} & \multirow{2}[4]{*}{TaoBao} & \multicolumn{5}{c}{Amazon} \\
\cmidrule{3-7}          &       & Women & Men   & Girls & Boys  & Baby \\
\midrule
    SC-GANs(-O) & 0.5947  & 0.5412 & 0.5318  & 0.5612  & 0.5042   & 0.5528   \\
    \midrule
    SC-GANs(-A) & 0.7119  & 0.6608  & 0.6463  & 0.6174  & 0.6007  & 0.5519  \\
    \midrule  
    SC-GANs(1G) & 0.6856 & 0.7187 & 0.7278 & 0.6373 & 	0.6815	& 0.6555 \\
    \midrule    
    SC-GANs & \textbf{0.8421} & \textbf{0.7899 } & \textbf{0.7906 } & \textbf{0.7778 } & \textbf{0.7574 } & \textbf{0.7434 }   \\    \bottomrule    
    \end{tabular}}
    \vspace{-3mm}

  \label{tab:ablation}
	
\end{table}

We then look into each component in SC-GANs, the orthogonal constraint, category-specific style transformations and adversarial learning strategy. 
We compare our full model with the following models.
\textbf{SC-GANs(-O)} doesn't have the orthogonal constraint on generators.
\textbf{SC-GANs(-A)} minimizes the Wasserstein distance 
without
using adversarial learning strategy \cite{Aude2016Stochastic}.
\textbf{SC-GANs(1G)} has only one transformation matrix for all categories.	
The experiments are conducted on the \emph{whole} Taobao datatset and Amazon dataset with 2\textperthousand ~ annotated relationships.    
             	
Performance comparison is shown in Table \ref{tab:ablation}. We notice that without the orthogonal constraint the performance is nearly equal to random guess.
A possible explanation is that the transformation matrix may try to map all the items into a tiny area, in which case it's hard to distinguish which are compatible. 
SC-GANs outperforms SC-GANs(-A), which implies that the adversarial learning strategy can better minimize the distances of distributions. 
Compared with SC-GANs(1G), the better performance of SC-GANs suggests that it's hard to learn one general transformation matrix for all the categories. Therefore, it's necessary to give different categories different transformation matrices.

\begin{figure}[t]
\centering
\includegraphics[width=0.97\linewidth]{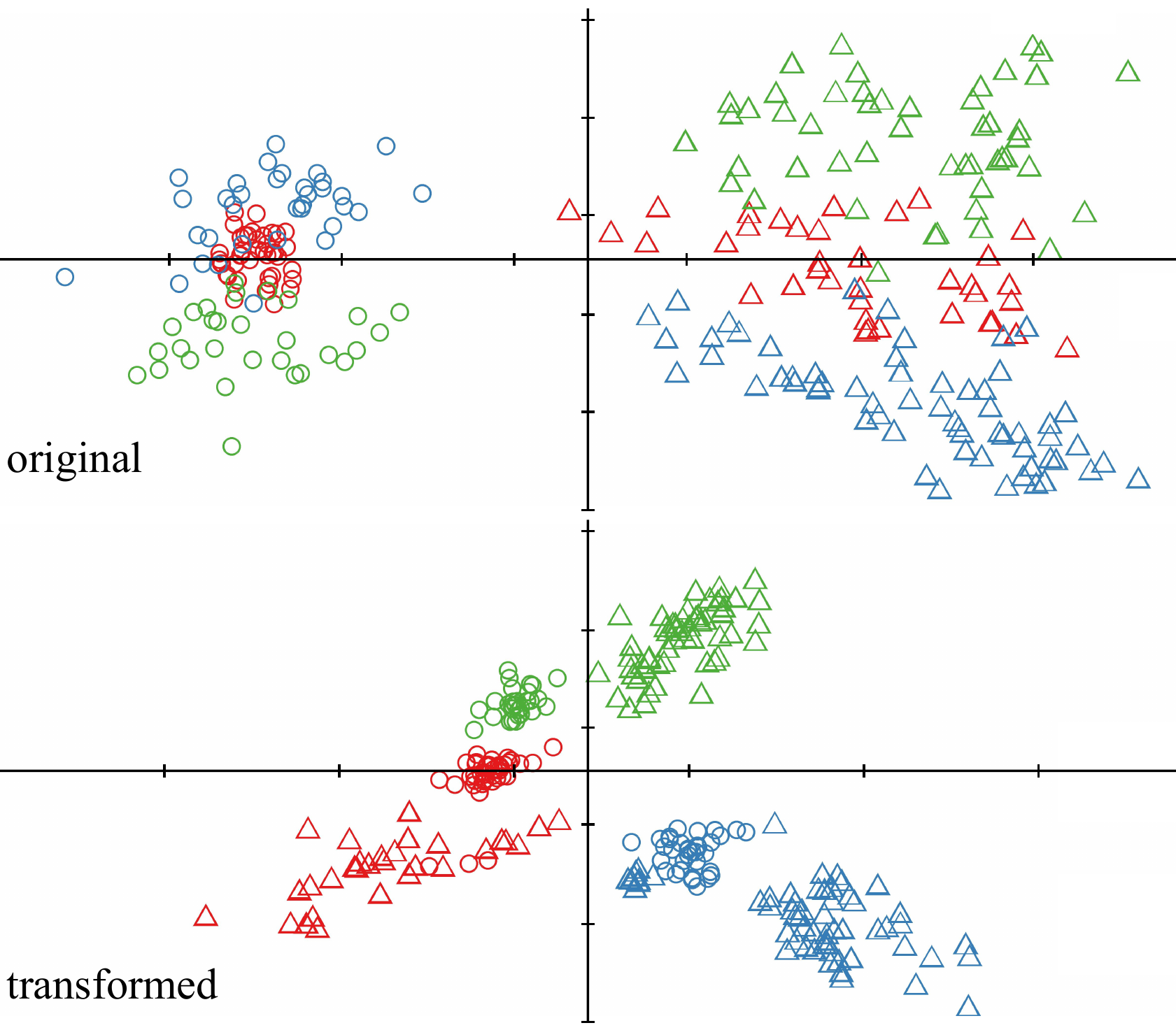}
\vspace{-4mm}
\caption{Visualization of the original and transformed distributions in the learned style space of tops and shoes. 
Circles represent shoes and triangles represent tops. 
The circles and triangles in same colors have same styles.
}
\label{fig:tsne}
\vspace{-6mm}
\end{figure}

\subsection{Visualization}
To intuitively illustrate 
that our proposed method can align the distributions and make compatible items with same styles close, 
we randomly select some tops and shoes from the Taobao dataset and show their original and transformed distributions in Figure \ref{fig:tsne}, with their raw image features and style vectors reduced to 2-D by PCA.
Circles represent shoes and triangles represent tops. 
The circles and triangles in same colors have same styles.
Red represents casual style, blue represents formal style and green represents sports style.
It can be seen the items in the original space cluster according to category and the compatible tops and shoes with same styles are distant. 
The distributions of tops and shoes have similar structures but different orientations and scales.
After transformed into the learned style space, the two distributions become close.
The items cluster according to style instead of category, which verifies the effectiveness of our method.

\section{Conclusions}
In this work, we first propose to consider the task of compatibility learning from item level to distribution level.
We find that aligning distributions of different categories can make compatible items with same styles close.
Achieving the alignment by minimizing the distance between distributions and also some annotated compatible items, we propose a semi-supervised method SC-GANs to learn compatibility across categories for clothing matching. 
In fact, the item-level distance minimization part in our work can be replaced with any supervised clothing matching method, which can be improved in the future work.

\section{Acknowledgments}
This work is supported by National Natural Science Foundation of China (61772528, 61871378) and National Key Research and Development Program (2016YFB1001000).

\bibliographystyle{IEEEbib}
\bibliography{icme2019template}

\begin{thebibliography}{10}

\bibitem{Mcauley2015Image}
Julian McAuley, Christopher Targett, Qinfeng Shi, and Anton Van Den~Hengel,
\newblock ``Image-based recommendations on styles and substitutes,''
\newblock in {\em SIGIR}, 2015.

\bibitem{He2017Learning}
Ruining He, Charles Packer, and Julian Mcauley,
\newblock ``Learning compatibility across categories for heterogeneous item
  recommendation,''
\newblock in {\em ICDM}, 2017.

\bibitem{Youn2016On}
H~Youn, L~Sutton, E~Smith, C~Moore, J.~F. Wilkins, I~Maddieson, W~Croft, and
  T~Bhattacharya,
\newblock ``On the universal structure of human lexical semantics,''
\newblock {\em Proceedings of the National Academy of Sciences of the United
  States of America}, 2016.

\bibitem{Zhang2017Earth}
Meng Zhang, Yang Liu, Huanbo Luan, and Maosong Sun,
\newblock ``Earth mover's distance minimization for unsupervised bilingual
  lexicon induction,''
\newblock in {\em EMNLP}, 2017.

\bibitem{Simonyan2014Very}
Karen Simonyan and Andrew Zisserman,
\newblock ``Very deep convolutional networks for large-scale image
  recognition,''
\newblock {\em ICLR}, 2015.

\bibitem{Veit2015Learning}
Andreas Veit, Balazs Kovacs, Sean Bell, Julian McAuley, Kavita Bala, and Serge
  Belongie,
\newblock ``Learning visual clothing style with heterogeneous dyadic
  co-occurrences,''
\newblock in {\em ICCV}, 2015.

\bibitem{liu2017deepstyle}
Qiang Liu, Shu Wu, and Liang Wang,
\newblock ``Deepstyle: Learning user preferences for visual recommendation,''
\newblock in {\em SIGIR}, 2017.

\bibitem{Rubner1998Metric}
Yossi Rubner, Carlo Tomasi, and L.~J. Guibas,
\newblock ``Metric for distributions with applications to image databases,''
\newblock in {\em ICCV}, 1998.

\bibitem{Goodfellow2014Generative}
Ian~J. Goodfellow, Jean Pouget-Abadie, Mehdi Mirza, Bing Xu, David
  Warde-Farley, Sherjil Ozair, Aaron Courville, and Yoshua Bengio,
\newblock ``Generative adversarial nets,''
\newblock in {\em NIPS}, 2014.

\bibitem{Arjovsky2017Wasserstein}
Martin Arjovsky, Soumith Chintala, and L{\'e}on Bottou,
\newblock ``Wasserstein gan,''
\newblock {\em ICML}, 2017.

\bibitem{Aude2016Stochastic}
Aude Genevay, Marco Cuturi, Gabriel Peyr{\'e}, and Francis Bach,
\newblock ``Stochastic optimization for large-scale optimal transport,''
\newblock in {\em NIPS}, 2016.

\end{thebibliography}

\end{document}